\documentclass[review]{elsarticle}
\usepackage{lineno,hyperref}
\usepackage{graphicx}
\usepackage{mathtools}
\usepackage{amsmath}
\usepackage{natbib}
\usepackage{mathrsfs}
\newcommand{\bra}[1]{\langle #1 \vert}

\newcommand{\ave}[1]{\langle #1 \rangle}
\newcommand{\beq}{\begin{equation}}
\newcommand{\eeq}{\end{equation}}
\newcommand{\beqa}{\begin{eqnarray}}
\newcommand{\eeqa}{\end{eqnarray}}
\newcommand{\nn}{\nonumber}
\newcommand{\rhoh}{\hat{\rho}}
\newcommand{\Tr}{\mbox{Tr}}
\newcommand{\hatU}{\hat{U}}
\newcommand{\ket}[1]{\vert #1 \rangle}
\newcommand{\hatH}{\hat{H}}

\newcommand{\hata}{\hat{a}}
\newcommand{\hatad}{\hat{a}^\dag}
\newcommand{\hatb}{\hat{b}}
\newcommand{\hatbd}{\hat{b}^\dag}
\newcommand{\hatc}{\hat{c}}
\newcommand{\hatcd}{\hat{c}^\dag}
\newcommand{\hatA}{\hat{A}}
\newcommand{\hatAd}{\hat{A}^\dag}
\newcommand{\hatB}{\hat{B}}
\newcommand{\hatBd}{\hat{B}^\dag}
\newcommand{\hatC}{\hat{C}}
\newcommand{\la}{\langle}
\newcommand{\ra}{\rangle}
\newcommand{\commentold}[1]{}

\modulolinenumbers[5]

\begin{document}

\begin{frontmatter}

\title{On quantum states generated from interacting oscillators}

\author{Haqi Ismael Shareef}
\ead{haqi.Ismail@garmian.edu.krd}
\address{Department of Physics, University of Kurdistan, P.O.Box 66177-15175, Sanandaj, Iran}
\address{Research center, Department of Physics, University of Garmian, Kalar, KRG, Iraq}

\author{Fardin Kheirandish}
\ead{f.kheirandish@uok.ac.ir}
\cortext[mycorrespondingauthor]{Corresponding author}
\address{Department of Physics, University of Kurdistan, P.O.Box 66177-15175, Sanandaj, Iran}
\begin{abstract}
In this paper, we study the quantum states generated from two and three linearly interacting quantum harmonic oscillators. We consider the possibility that one of the oscillators be under the influence of a classical external source and obtain the total and reduced density matrices related to the system. We show how the problem can be generalized to $n$-linearly interacting oscillators straightforwardly.
\end{abstract}

\end{frontmatter}

\section{Introduction}\label{Introduction}
Early work on the generation of nonclassical states was based on the motional states of trapped ions in the late 90s \cite{d0,d1,d2,d3,d4,d5,d6,d7}. In recent years, the generation of Fock or number states and their superpositions have drawn much attention \cite{r0,r1, r2, r3, r4, r5}.
One of the most important subjects in quantum optics is the generation and manipulation of the nonclassical states \cite{b1}. An important class of nonclassical states is Fock states that play an important role in quantum information processing \cite{b2,b3,b4}. Some important nonclassical states are made by superpositions of Fock states like coherent states \cite{b5}, Cat states \cite{b6}, displaced Fock states \cite{b7}, squeezed states \cite{b8}, and squeezed coherent states \cite{b9}. Squeezed states are the main ingredients in high-precision measurements, and the nonclassical states with a binomial distribution have applications in quantum error correction \cite{b10,b11}. Fock states are eigenstates of the Hamiltonian of a quantum harmonic oscillator. The preparation of the harmonic oscillator in a definite Fock state or a specific superposition of Fock states may not be a trivial task. Usually, external classical sources are applied to the oscillator to generate a specific state, but this may lead to an undesired thermal state which can be avoided by introducing another auxiliary system to work between the external sources and the main system \cite{c17,c18}. The auxiliary-system technique has been used for the preparation of Fock states in an optical cavity with Rydberg atoms \cite{c19,c20,c21,c22,c23}. In \cite{c24} a classical field has been applied to the oscillator directly without using the auxiliary systems.

Here, in the framework of Heisenberg equations and Bogoliubov transformations, we consider two and three linearly interacting quantum oscillators. In Sec. (\ref{Secii}), for two linearly interacting oscillators denoted by $a,\,b$-oscillators, where also an external classical source is applied to the $b$-oscillator, the generated binomial states, the energy exchange between the oscillators, and matrix elements of the reduced density matrix of the $a$-oscillator are obtained and discussed. In Sec. (\ref{Seciii}), we will generalize the approach to three linearly interacting quantum oscillators denoted by $a,\,b,\,c$-oscillators. In the following, we explain how the problem can be generalized to $n$-linearly interacting oscillators.
\section{Two interacting oscillators}\label{Secii}
The Hamiltonian that we are considering in this section is the Hamiltonian of two interacting bosonic modes (Fig. 1) with an arbitrary time-dependent coupling $g(t)$ given in Eq. (\ref{H1}). Here, the coupling function between the oscillators is assumed to be time-dependent to keep generality and have control on the coupling strength. For example, as a particular case, $g(t)$ can be considered as a switching function that is on during a specific time interval $\tau$ and is off for $t>\tau$
\beq\label{switch}
g(t)=\left\{
       \begin{array}{ll}
         g_0, & t\leq \tau, \\
         0, & t> \tau,
       \end{array}
     \right.
\eeq
where $g_0$ is a constant.
\begin{figure}[h]
  \centering
  \includegraphics[width=.5\textwidth]{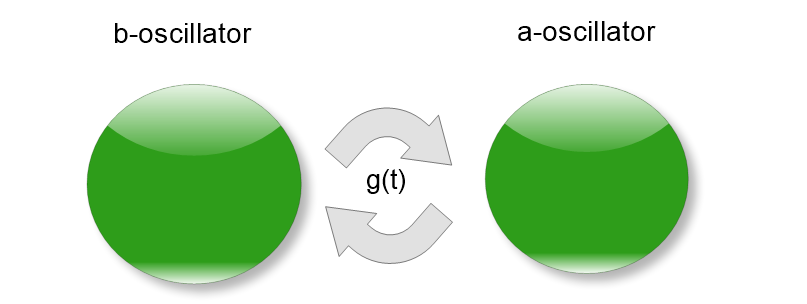}
  \caption{(color online) Two interacting oscillators with a time-dependent coupling function $g(t)$. }\label{Fig1}
\end{figure}
The total Hamiltonian is
\beq\label{H1}
\hatH (t)=\underbrace{\hbar\omega_0\,\hatad \hata}_{\hat{H}_a}+\underbrace{\hbar\Omega\,\hatbd\hatb}_{\hat{H}_b}+\underbrace{\hbar g(t)\,(\hata\hatbd+\hatad\hatb)}_{\hat{H}_{ab}(t)},
\eeq
which can be written in matrix form
\beq\label{Matrix2}
\hatH (t)=\hbar\,\left(
                   \begin{array}{cc}
                     \hata^\dag & \hatb^\dag \\
                   \end{array}
                 \right)\left(
                          \begin{array}{cc}
                            \omega_0 & g(t) \\
                            g(t) & \Omega \\
                          \end{array}
                        \right)\left(
                                 \begin{array}{c}
                                   \hata \\
                                   \hatb \\
                                 \end{array}
                               \right).
\eeq
Let the combined system be initially prepared in an arbitrary state $|\psi(0)\ra$, to find the time-evolution of the state $|\psi(t)\ra=\hat{U}(t)\,|\psi(0)\ra$, we need to find the unitary evolution operator $\hat{U}(t)$. For this purpose, we diagonalize the Hamiltonian Eq. (\ref{H1}) using the following time-dependent Bogoliubov transformations
\beqa\label{H2}
 \hata &=& \cos\theta(t)\,\hatA+\sin\theta(t)\,\hatB,\\
 \hatb &=& -\sin\theta(t)\,\hatA+\cos\theta(t)\,\hatB,\label{H2-2}
\eeqa
where $[\hatA,\hatB]=[\hatA,\hatBd]=0$ and $\theta(t)$ is a time-dependent parameter to be defined suitably. By inserting $\hata\,(\hatad)$ and $\hatb \,(\hatbd)$ from Eqs. (\ref{H2},\ref{H2-2}) into Eq. (\ref{H1}), the Hamiltonian is decomposed into two independent oscillators as follows,
\beq\label{H3}
\hatH=\hbar\lambda_A (t)\,\hatAd\hatA+\hbar\lambda_B (t)\,\hatBd\hatB,
\eeq
where the new frequencies corresponding to $A$ and $B$-oscillators are defined respectively by
\beqa\label{H4}
&& \lambda_A (t) = \bar{\omega}-\sqrt{g^2 (t)+\triangle^2/4},\\\label{H4-2}
&& \lambda_B (t) = \bar{\omega}+\sqrt{g^2 (t)+\triangle^2/4},\\
&& \tan(2\theta(t)) = 2 g(t)/\triangle.
\eeqa
In Eqs. (\ref{H4},\ref{H4-2}), $\bar{\omega}=(\Omega+\omega_0)/2 $ is the average frequency and $\triangle=\Omega-\omega_0\geq 0$ is the detuning. For later convenience, we write $\theta$ instead of $\theta(t)$, and its time dependence is implicitly assumed. In the case of resonance $\Omega=\omega_0$, we have $\triangle=0$ and Eqs. (\ref{H4},\ref{H4-2}) are reduced to
\beqa\label{H5}
\lambda_A (t) &=& \omega_0-|g(t)|,\\
\lambda_B (t) &=& \omega_0+|g(t)|,
\eeqa
and $|\theta|\rightarrow\pi/4$.
\subsection{Energy exchange}\label{Seciii}
In Hamiltonian Eq. (\ref{H1}), let the $a$-oscillator be prepared in the number state $\ket{n}_a$ and the $b$-oscillator be prepared in the vacuum state $\ket{0}_b$. Note that the vacuum state of the Hamiltonian Eq. (\ref{H3}) is $\ket{0}_A\otimes\ket{0}_B$ with zero energy, and the state $\ket{0}_a\otimes\ket{0}_b$ is the vacuum state of the same Hamiltonian expressed as Eq. (\ref{H1}). Therefore, due to the uniqueness of the vacuum state, we have $\ket{0}_a\otimes\ket{0}_b\equiv \ket{0}_A\otimes\ket{0}_B$. Accordingly, the initial state can be written as
\beqa\label{H6}
\ket{\psi(0)} &=& \ket{n}_a\otimes\ket{0}_b,\nn\\
              &=& \frac{(\hatad)^n}{\sqrt{n!}}\,\ket{0}_a\otimes\ket{0}_b,\nn\\
              &=& \frac{(\cos(\theta)\hatAd+\sin(\theta)\hatBd)^n}{\sqrt{n!}}\,\ket{0}_A\otimes\ket{0}_B,\nn\\
              &=& \sum_{r=0}^n \sqrt{\binom{n}{r}}(\cos(\theta))^{n-r} (\sin(\theta))^r\,\ket{n-r}_A\otimes\ket{r}_B,\nn\\
              &=& (\cos(\theta))^n \sum_{r=0}^n \sqrt{\binom{n}{r}}(\tan(\theta))^r\,\ket{n-r}_A\otimes\ket{r}_B,
\eeqa
where $\ket{m}_{A(B)}$ belongs to the Hilbert space of $A(B)$-oscillator, respectively. The unitary evolution operator corresponding to Hamiltonian Eq. (\ref{H3}) is the tensor product of the unitary operators corresponding to $A(B)$-oscillators
\beqa\label{unitary}
\hat{U}(t)=&& \hat{U}_A (t)\otimes\hat{U}_B (t),\nn\\
=&& e^{-if_A (t)\,\hatAd\,\hatA}\otimes e^{-if_B (t)\,\hatBd\,\hatB},
\eeqa
where for notational simplicity we have defined
\beq\label{fab}
f_{A(B)} (t)=\int_0^t dt'\,\lambda_{A(B)} (t').
\eeq
By applying $\hat{U}(t)$ on the initial state Eq. (\ref{H6}), we find the evolved state as
\beqa\label{H7}
\ket{\psi(t)}=[e^{-if_A (t)}\cos(\theta)]^{n}\sum_{r=0}^n \sqrt{\binom{n}{r}}(\tan(\theta))^r\,
e^{ir\,\delta f(t)}\ket{n-r}_A\otimes\ket{r}_B,
\eeqa
where for notational simplicity we have defined
\beqa\label{deltaf}
\delta f(t) &=& f_A (t)-f_B (t)=-2\int_0^t dt'\,\sqrt{g^2(t')+\triangle^2}.
\eeqa
The state $\ket{\psi(t)}$ in Eq. (\ref{H7}) is expressed in $A(B)$-oscillator number states or Fock states. By using the inverse of the Bogoliubov transformations Eq. (\ref{H2})
\beqa\label{Inverse}
 \hatA &=& \cos\theta(t)\,\hata-\sin\theta(t)\,\hatb,\\
 \hatB &=& \sin\theta(t)\,\hata+\cos\theta(t)\,\hatb,\label{Inverse-2}
\eeqa
and following the same process we used in Eq. (\ref{H6}), the state $\ket{\psi(t)}$ can be represented in the initial $a(b)$-oscillator states as
\beqa\label{H8}
\ket{\psi(t)} =&& (\cos(\theta))^{2n}\sum_{r=0}^n \sum_{p=0}^{n-r}\sum_{q=0}^r\, \frac{(-1)^p\,\sqrt{n!\,(n-p-q)!\,(p+q)!}}{p!\,q!\,(n-r-p)!\,(r-q)!}\nn\\
&& \times(\tan(\theta))^{2r+p-q}e^{-i(n-r)f_A (t)}\,e^{-ir f_B(t)}\,\ket{n-(p+q)}_a\otimes\ket{p+q}_b.\nn\\
\eeqa
\subsection{Binomial states}
As an example, let us set $n=1$, in Eq. (\ref{H8}) then
\beq\label{H9}
\ket{\psi(t)}=C_{10}(t)\,\ket{1}_a\otimes\ket{0}_b+C_{01}(t)\,\ket{0}_a\otimes\ket{1}_b,
\eeq
where
\beqa\label{c10}
|C_{10}(t)|^2 &=& \frac{4g^2 (t)\,\cos^2[\delta f(t)/2]+\triangle^2}{4g^2(t)+\triangle^2}=q(t),\\
|C_{01}(t)|^2 &=& \frac{4g^2 (t)\,\sin^2[\delta f(t)/2]}{4g^2(t)+\triangle^2}=p(t),\label{c01}
\eeqa
and $q(t)+p(t)=1$. Also, by setting $n=2$ in the same equation, we find
\beq\label{H11}
\ket{\psi(t)}=C_{20}(t)\,\ket{2}_a\otimes\ket{0}_b+C_{11}(t)\,\ket{1}_a\otimes\ket{1}_b+
C_{02}(t)\,\ket{0}_a\otimes\ket{2}_b,
\eeq
\beqa\label{H9}
|C_{20}(t)|^2 &=& \left[\frac{4g^2(t)\,\cos^2[\delta f(t)/2]+\triangle^2}{4g^2(t)+\triangle^2}\right]^2=q^2(t),\\
|C_{02}(t)|^2 &=& \left[\frac{4g^2(t)\,\sin^2[\delta f(t)/2]}{4g^2(t)+\triangle^2}\right]^2=p^2(t),\\
|C_{11}(t)|^2 &=& \frac{4g^2 (t)\Big[2g^2(t)+\triangle^2-2g^2(t)\cos^2(\delta f(t))-\triangle^2
\cos(\delta f(t))\Big]}{(4g^2 (t)+\triangle^2)^2},\nn\\
              &=& 2 q(t) p(t).
\eeqa
In general, for the initial state $\ket{\psi(0)}=\ket{n}_a\otimes\ket{0}_b$, one finds the evolved state as a binomial state given by
\beq\label{psit}
\ket{\psi(t)}=\sum_{s=0}^n C_{n-s,s}(t)\,\ket{n-s}_a\otimes\ket{s}_b,
\eeq
where the coefficients are given by
\beqa\label{ens}
&& C_{n-s,s}(t)= \nn\\
&& e^{-inf_a (t)}(\cos(\theta))^{2n}\sqrt{\binom{n}{s}}(1+\tan^2 (\theta)\,e^{i\delta f (t)})^{n-s}\,(\tan(\theta))^s (e^{i\delta f(t)}-1)^s.\nn\\
\eeqa
Therefore, the probability that the system be found in the state $\ket{n-s}_a\otimes\ket{s}_b$ at time $t$ is
\beqa\label{H12}
|C_{n-s,s}(t)|^2 =&& \binom{n}{s}\,\Bigg[\frac{4g^2(t)\,
\cos^2[\delta f(t)/2]+\triangle^2}{4g^2(t)+\triangle^2}\Bigg]^{n-s}\,
\Bigg[\frac{4g^2(t)\,\sin^2[\delta f(t)/2]}{4g^2(t)+\triangle^2}\Bigg]^{s},\nn\\
=&& \binom{n}{s}\,(q(t))^{n-s} (p(t)^s.
\eeqa
which is a binomial distribution that reminiscences a random walk with the time-dependent right ($p(t)$) and left ($q(t)$) probabilities given in Eqs. (\ref{c10}, \ref{c01}), respectively.

On the resonance ($\Delta=0$), we can set $\theta=\pi/4$, and the Eqs. (\ref{psit},\ref{H12}) become
\beqa\label{ensr}
\ket{\psi(t)}_{reso.} =&& e^{-i n (f_A(t)+f_B(t))/2}\,(\cos(\delta f(t)/2))^n\nn\\
&& \times\sum_{s=0}^n \sqrt{\binom{n}{s}}\,(i\tan(\delta f(t)/2))^s \,\ket{n-s}_a\otimes\ket{s}_b,
\eeqa
and
\beqa\label{ensr2}
|C_{n-s,s}(t)|_{reso.}^2 = \binom{n}{s}\,\big(\cos^2 (\delta f(t)/2)\big)^{n-s}\,\big(\sin^2 (\delta f(t)/2)\big)^{s},
\eeqa
respectively. Note that on the resonance we have
\beq\label{remark1}
\delta f(t)=-2 \int_0^t dt'\,|g(t')|=-2 G(t).
\eeq
For a constant coupling $g(t)=g_0$, we have $\delta f(t)=-2 g_0 t$, and at the time $t_0=\pi/(2 g_0)$, we have $\cos(\delta f(t_0)/2)=0$, and from Eq. (\ref{ensr2}) we find
\beq\label{remark1}
|C_{n-s,s}(t_0)|_{reso.}=\left\{
                           \begin{array}{ll}
                             0, & 0\leq s\leq n-1, \\
                             1, & s=n.
                           \end{array}
                         \right.
\eeq
Therefore, by using Eq. (\ref{psit}) we deduce that at $t=t_0$ the initial state $\ket{n}_a\otimes\ket{0}_b$ has transformed to the state $\ket{0}_a\otimes\ket{n}_b$.
\subsection{The reduced density matrix}\label{DM}
From Eq. (\ref{psit}), the total density matrix at time $t$ is
\beq\label{H13}
\rhoh (t)=\ket{\psi(t)}\bra{\psi(t)}=\sum_{s=0}^n \sum_{s'=0}^n C_{n-s,s}\,\bar{C}_{n-s',s'}\ket{n-s}_{a\,a}\bra{n-s'}\otimes\ket{s}_{b\,b}\bra{s'},
\eeq
and the reduced density matrix corresponding to the $a$-mode can be obtained by tracing out the $b$-mode degrees of freedom as
\beq\label{reda}
\rhoh_a (t) =\Tr_b \,[\rhoh (t)]= \sum_{s=0}^n |C_{n-s,s}|^2\,\ket{n-s}_{a\,a}\bra{n-s},
\eeq
where $\Tr_b$ is the partial trace over $b$-mode degrees of freedom. Having the reduced density matrix $\rhoh_a (t)$ we can find the average energy of the $a$-mode oscillator at time $t$ as
\beqa\label{ener}
\ave{\hat{H}_a} &=& \Tr[\rhoh_a (t)\,\hat{H}_a]=\hbar\omega_0\,\sum_{s=0}^n |C_{n-s,s}|^2\,(n-s),\nn\\
&=& \hbar\omega_0\,\sum_{s=0}^n \binom{n}{s}\,(q(t))^{n-s} (p(t))^s\,(n-s)\nn\\
&=& \hbar\omega_0\,\Big(n-p\frac{\partial}{\partial p}\Big)(p+q)^n,\nn\\
&=& n\hbar\omega_0 q(t),
\eeqa
where on resonance $q(t)=\cos^2(\delta f(t)/2)=\cos^2(G(t))$.
\subsection{The initial state is a coherent state}
Let the initial state of the combined system be given by the separable state $\ket{\varphi(0)}=\ket{\alpha}_a\otimes\ket{0}_b$, where $\ket{\alpha}_a$ is a coherent state of the $a$-oscillator. The state of the system at time $t$ is
\beqa\label{coh1}
\ket{\varphi(t)} &=& \hatU (t) \ket{\alpha}_a\otimes\ket{0}_b=e^{-|\alpha|^2/2}\sum_{n=0}^\infty \frac{\alpha^n}{\sqrt{n!}}\underbrace{\hatU (t) \ket{n}_a\otimes\ket{0}_b}_{\ket{\psi (t)}\,in\,Eq. (\ref{psit})},\nn\\
                 &=& e^{-|\alpha|^2/2}\sum_{n=0}^\infty \frac{\alpha^n}{\sqrt{n!}}\sum_{s=0}^n C_{n-s,s}(t)\,\ket{n-s}_a\otimes\ket{s}_b,\nn\\
                 &=& e^{-|\alpha|^2/2}\sum_{n=0}^\infty \frac{\Big(\alpha\,\cos^2(\theta)[e^{-if_A (t)}+\tan^2 (\theta)\,e^{-if_B (t)}]\Big)^n}{\sqrt{n!}}\nn\\
                 && \times\sum_{s=0}^n \sqrt{\binom{n}{s}}\Big(\frac{\tan(\theta) [e^{i\delta f(t)}-1]}{1+\tan^2(\theta)\,e^{i\delta f(t)}}\Big)^s\,\ket{n-s}_a\otimes\ket{s}_b,\nn\\
\eeqa
and the probability of finding the oscillators in the state $|n-s\ra_a\otimes|s\ra_b$ is
\beqa
\underbrace{\frac{e^{-|\alpha|^2}|\alpha\cos^2 (\theta)[1+\tan^2(\theta)e^{i\delta f(t)}]|^{2n}}{n!}}_{Poisson}\underbrace{\binom{n}{s}\Bigg|\frac{\tan(\theta)[e^{i\delta f(t)}-1]}{1+\tan^2(\theta)e^{i\delta f(t)}}\Bigg|^2}_{binomial},
\eeqa
which is the product of a Poisson and a binomial distribution.

On the resonance, Eq. (\ref{coh1}) becomes
\beqa\label{coh1reso}
&& \ket{\varphi(t)}=\nn\\
&& e^{-\frac{|\alpha|^2}{2}}\sum_{n=0}^\infty \frac{\big(\alpha\,e^{-i \omega_0 t}\,\cos(G(t))\big)^n}{\sqrt{n!}}\sum_{s=0}^n \sqrt{\binom{n}{s}}\big(-i\tan(G(t))\big)^s\,\ket{n-s}_a\otimes\ket{s}_b,\nn\\
\eeqa
and in this case, one finds the reduced density matrix for the $a$-oscillator as
\beqa\label{redc}
&& \rhoh_a (t) = \Tr_b \big[\ket{\varphi(t)}\bra{\varphi(t)}\big]=\nn\\
&&             e^{-|\alpha|^2/2}\sum_{n,n'=0}^\infty\sum_{s=0}^n
            \frac{\big(\alpha\,e^{-i \omega_0 t}\,\cos(G(t))\big)^n\,\big(\bar{\alpha}\,e^{i \omega_0 t}\,\cos(G(t))\big)^{n'}}{\sqrt{n! n'!}}\nn\\
&& \times \sqrt{\binom{n}{s}\binom{n'}{s}}
            \big(\tan^2 (G(t))\big)^s\,\ket{n-s}_a\otimes\ket{s}_b.\nn\\
\eeqa
By making use of Eq. (\ref{redc}), the average energy of $a$-oscillator is obtained as
\beqa\label{avee}
\ave{\hatH_a}=\Tr [\rhoh_a (t)\,\hatH_a]=\hbar\omega_0 |\alpha|^2 \cos^2 (G(t)).
\eeqa
For a constant coupling, we have $g(t)=g_0$, ($G(t)=g_0 t$), therefore, the $a$-oscillator loses all its energy at time $t_0=\pi/2g_0$ which is transferred to the $b$-oscillator. By using Eq. (\ref{coh1}), one finds the state of the system at time $t=t_0$ as $\ket{\varphi(t_0)}=\ket{0}_a\otimes\ket{-i\alpha e^{-i\omega_0 t_0}}_b$. If we chose $g_0=\omega_0/3$, then the final state at time $t_0$ will be $\ket{0}_a\otimes\ket{\alpha}_b$, that is, the coherent state has transferred to the $b$-oscillator.
\subsection{A scheme for dissipation}
For a decaying coupling in time like $g(t)=g_0 e^{-t/\tau}$, we have
\beq\label{diss1}
G(t)=\int_0^t dt'\,g(t')=g_0 \tau (1-e^{-t/\tau}),
\eeq
and both Eqs. (\ref{ener},\ref{avee}) behave like $\cos^2 (G(t))$. In Eqs. (\ref{ener},\ref{avee}) the scaled energies $\ave{\hatH_a}/n\hbar\omega_0$ and $\ave{\hatH_a}/\hbar\omega_0 |\alpha|^2$ are the same and behave as $E(t)=\cos^2 (G(t))$. The scaled energy is depicted in Fig.1 in terms of the dimensionless variable $x=g_0 t$ for $g_0 \tau=\pi/2$.
\begin{figure}[h]
  \centering
  \includegraphics[width=.5\textwidth]{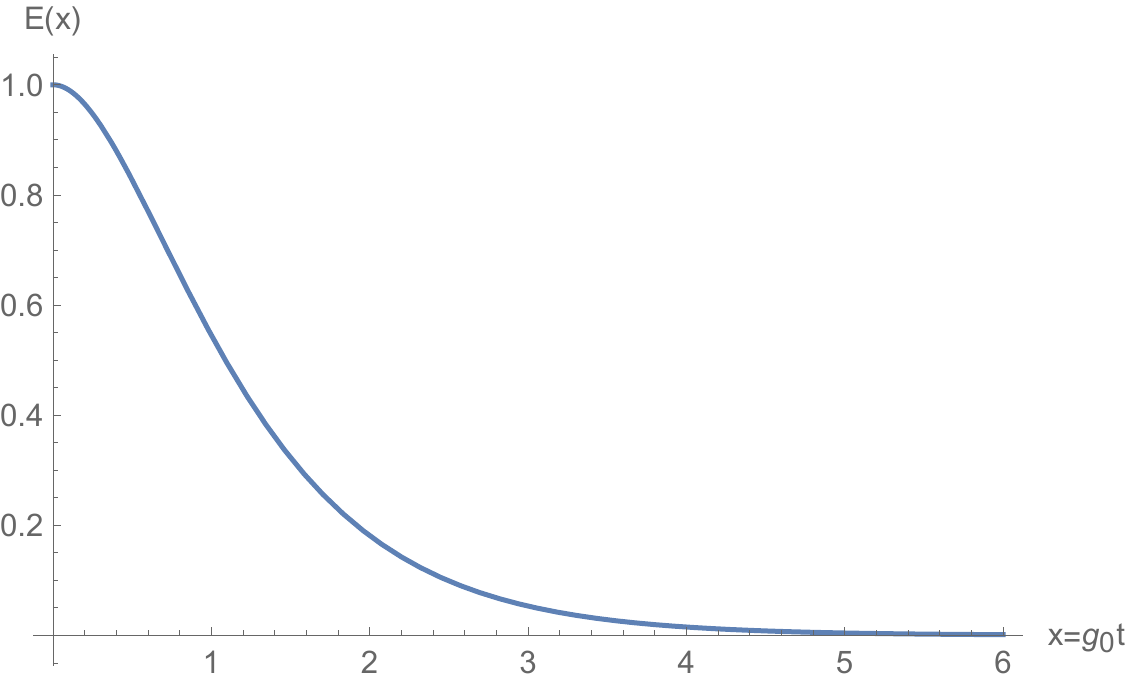}
  \caption{(color online) The scaled energy $E(x)=\cos^2[\frac{\pi}{2}(1-e^{-\frac{2x}{\pi}})]$ of the $a$-oscillator in terms of the dimensionless variable $x=g_0 t$ for $g_0 \tau=\pi/2$.}\label{Fig2}
\end{figure}
From Fig.2, one can easily deduce that the flow of energy from the $a$-oscillator to the $b$-oscillator is rectified, and the energy can not return to the $a$-oscillator due to the vanishing behavior of the coupling function $g(t)$ in time.
\subsection{External classical source}
In this section, we consider the effect of an external time-dependent classical source acting linearly on the $b$-oscillator. Here, the main subsystem is the $a$-oscillator. The total Hamiltonian in this case is
\beq\label{ext1}
\hatH^{(k)} (t)=\underbrace{\hbar\omega_0\,\hatad \hata}_{\hat{H}_a}+\underbrace{\hbar\Omega\,\hatbd\hatb}_{\hat{H}_b}+\underbrace{\hbar g(t)\,(\hata\hatbd+\hatad\hatb)}_{\hat{H}_{ab}(t)}+\underbrace{\hbar k(t)\hatbd+\hbar\bar{k}(t)\hatb}_{\hatH_{bs}}.
\eeq
By using the same Bogoliubov transformations Eqs. (\ref{H2},\ref{H2-2}), the Hamiltonian can be transformed to the following form
\beqa\label{ext2}
\hatH^{(k)} (t)=&& \underbrace{\hbar\lambda_A (t)\,\hatAd\hatA-\hbar\sin\theta\,(k(t)\hatAd+\bar{k}(t)\hatA)}_{\hatH_A (t)}\nn\\
&& +\underbrace{\hbar\lambda_B (t)\hatBd\hatB+\hbar\cos\theta\,(k(t)\hatBd+\bar{k}(t)\hatB)}_{\hatH_B (t)},
\eeqa
and the time-evolution operator can be written immediately as
\beq\label{ext3}
\hatU^{(k)} (t)=\hatU_A (t)\otimes\hatU_B (t),
\eeq
where
\beq\label{ext4}
i\hbar\,\frac{d\hatU_{A(B)} (t)}{dt}=\hatH_{A(B)} (t)\,\hatU_{A(B)} (t).
\eeq
Our aim is to find the unitary operator $\hatU_A (t)$. Note that the Hamiltonian $\hatH_B$ can be obtained from the Hamiltonian $\hatH_A $ using the transformations $\sin\theta\rightarrow -\cos\theta,\,A\rightarrow B$, therefore, one can obtain $\hatU_B (t)$ from $\hatU_A (t)$ using the same transformations. To find $\hatU_A (t)$ let us make the ansatz
\beq\label{ext5}
\hatU_A (t)=e^{-i f_0^A (t)/\hbar}\,e^{-i f_1^A (t)\hatAd\hatA/\hbar}\,e^{-i f_2^A (t)\hatAd/\hbar}\,e^{-i \bar{f}_2^A (t)\hatA/\hbar}.
\eeq
By inserting $\hatU_A (t)$ into Eq. (\ref{ext4}) we will find the unknown functions $f_0 (t)$, $f_1 (t)$ and $f_2 (t)$ as
\beqa\label{ext6}
&& f_1^A (t)=\hbar\int_0^t dt'\,\lambda_A (t')=\hbar\,f_A (t),\nn\\
&& f_2^A (t)= -\hbar\,\int_0^t dt'\,\sin\theta(t')\,k(t')\,e^{i f_A (t')},\nn\\
&& f_0^A (t)= -\frac{i}{2\hbar}\,|f_2^A (t)|^2.
\eeqa
In the case of resonance, we can set $\theta=\pi/4$, and also by assuming a time-independent coupling constant $g(t)=g_0$, we find the following explicit forms for the unknown functions
\beqa\label{ext7}
&& f_1^A (t)=\hbar(\omega_0-g_0) t,\nn\\
&& f_2^A (t)= -\frac{\hbar}{\sqrt{2}}\,\int_0^t dt'\,k(t')\,e^{i (\omega_0-g_0) t'},\nn\\
&& f_0^A (t)= \frac{-i}{4\hbar}\Big|\int_0^t dt'\,k(t')\,e^{i(\omega_0-g_0)t'}\Big|^2.
\eeqa
Similarly, we can assume the following form for $\hatU_B$
\beq\label{ext8}
\hatU_B (t)=e^{-i f_0^B (t)/\hbar}\,e^{-i f_1^B (t)\hatBd\hatB/\hbar}\,e^{-i f_2^B (t)\hatBd/\hbar}\,e^{-i \bar{f}_2^B (t)\hatB/\hbar},
\eeq
and find the unknown functions from the previously mentioned transformations $\sin\theta\rightarrow -\cos\theta,\,A\rightarrow B$ as
\beqa\label{ext9}
&& f_1^B (t)=\hbar\int_0^t dt'\,\lambda_B (t')=\hbar\,f_B (t),\nn\\
&& f_2^B (t)= \hbar\,\int_0^t dt'\,\cos\theta(t')\,k(t')\,e^{i f_B (t')},\nn\\
&& f_0^B (t)= -\frac{i}{2\hbar}\,|f_2^B (t)|^2.
\eeqa
%
\subsection{The evolution of the initial state}
let us assume that initially (at $t=0$), each oscillator is prepared in its ground state, then the initial state of the combined system is
\beq\label{gr1}
\ket{\psi(0)}=\ket{0}_a\otimes\ket{0}_b\equiv\ket{0}_A\otimes\ket{0}_B=\ket{0}.
\eeq
By applying the time-evolution operator Eq. (\ref{ext3}) on the initial state, we will find the evolved state as
\beqa\label{gr2}
\ket{\psi(t)} &=& \hatU_a \ket{0}_A\otimes\hatU_B \ket{0}_B,\nn\\
              &=& e^{-i(f_0^A (t)+f_0^B (t))/\hbar}\,e^{(|f_2^A (t)|^2+|f_2^B (t)|^2)/2\hbar^2}\nn\\
              && \times\Big |{-\frac{i}{\hbar}f_2^A (t)\,e^{-i f_1^A (t)/\hbar}}\Big\ra_A\otimes
              \Big |{-\frac{i}{\hbar}f_2^B (t)\,e^{-i f_1^B (t)/\hbar}}\Big\ra_B.
\eeqa
For notational simplicity, we define the following time-dependent functions
\beq\label{gr3}
\mu_{A(B)} (t):=-\frac{i}{\hbar}f_2^{A(B)} (t)\,e^{-i f_1^{A(B)} (t)/\hbar}.
\eeq
Now using the definition of a coherent state and Bogoliubov transformations, Eq. (\ref{gr2}) can be rewritten as
\beqa\label{gr4}
\ket{\psi(t)} &=& e^{-i(f_0^A (t)+f_0^B (t))/\hbar}\,\sum_{n,m=0}^\infty\frac{(\mu_A)^n}{n!}\frac{(\mu_B)^m}{m!}\,\ket{n}_A\otimes\ket{m}_B,\nn\\
              &=&  e^{-i(f_0^A (t)+f_0^B (t))/\hbar}\nn\\
              && \times\sum_{n,m=0}^\infty\frac{(\mu_A)^n}{n!}\frac{(\mu_B)^m}{m!}\,(\cos\theta\,\hatad-\sin\theta\,\hatbd)^n
              (\sin\theta\,\hatad+\cos\theta\,\hatbd)^m\,\ket{0},\nn\\
              &=& e^{-i(f_0^A (t)+f_0^B (t))/\hbar}\,e^{(\mu_A \cos\theta+\mu_B\sin\theta)\hatad}e^{(\mu_B \cos\theta-\mu_A\sin\theta)\hatbd}\,\ket{0},\nn\\
              &=& e^{-i(f_0^A (t)+f_0^B (t))/\hbar}\nn\\
              && \times\sum_{n,m=0}^\infty \frac{(\mu_A \cos\theta+\mu_B\sin\theta)^n}{\sqrt{n!}}
              \frac{(\mu_B \cos\theta-\mu_A\sin\theta)^m}{\sqrt{m!}}\,\ket{n}_a\otimes\ket{m}_b.\nn\\
\eeqa
Therefore, the probability that at time $t$ we find the combined system in the product state $\ket{n}_a\otimes\ket{m}_b$ is
\beq\label{gr5}
|C_{n,m}^{(k)} (t)|^2=\frac{|\mu_A \cos\theta+\mu_B\sin\theta|^{2n}}{n!}\,\frac{|\mu_B \cos\theta-\mu_A\sin\theta|^{2m}}{m!}\,e^{\frac{2}{\hbar} \mbox{Im}(f_0^A+f_0^B)},
\eeq
where $\mbox{Im}(g)$ gives the imaginary part of a complex function $g$. By using Eqs. (\ref{ext6},\ref{ext9}), one easily finds
\beq\label{gr6}
\sum_{n=0}^\infty \sum_{m=0}^\infty |C_{n,m}^{(k)} (t)|^2=1,
\eeq
as expected. From Eq. (\ref{gr4}) one also finds $\ket{\psi(t)}$ as a separable state of two coherent states
\beq\label{gr7}
\ket{\psi(t)}=\ket{\mu_A \cos\theta+\mu_B\sin\theta}_{a}\otimes\ket{\mu_B\cos\theta-\mu_A\sin\theta}_{b}.
\eeq
In the case of resonance ($\theta=\pi/4$) and a time-independent coupling function, we have
\beq\label{gr8}
\ket{\psi(t)}=\ket{\alpha_A}_{a}\otimes\ket{\alpha_B}_{b},
\eeq
where
\beqa\label{gr9}
\alpha_A &=& \frac{\mu_B+\mu_A}{\sqrt{2}}=-\frac{1}{2}\int_0^t dt'\,k(t')e^{-i\omega_0 (t-t')}\sin g_0 (t-t'),\\
\alpha_B &=& \frac{\mu_B-\mu_A}{\sqrt{2}}=-\frac{i}{2}\int_0^t dt'\,k(t')e^{-i\omega_0 (t-t')}\cos g_0 (t-t').\label{gr9-2}
\eeqa
%
\subsection{The reduced density matrix $\rhoh_a (t)$}
The elements of the reduced density matrix $\hat{\rho}_a (t)$ corresponding to the main system ($a$-oscillator) initially prepared in the ground state, can be obtained from the state Eq. (\ref{gr4}) as
\beqa\label{Red1}
{}_a\bra{p}\rhoh_a (t)\ket{q}_a=&& \Tr_b \left[\ket{\psi(t)}\bra{\psi(t)}\right],\nn\\
=&& e^{\frac{2}{\hbar}\mbox{Im}[f_0^A(t)+f_0^B(t)]}\,e^{|\mu_B\cos\theta-\mu_A\sin\theta|^2}\nn\\
&& \times\,\frac{(\mu_B\cos\theta+\mu_A\sin\theta)^p}{\sqrt{p!}}
\frac{(\bar{\mu}_B\cos\theta+\bar{\mu}_A\sin\theta)^q}{\sqrt{q!}}.
\eeqa
The probability that the $a$-oscillator be found in the state $\ket{n}_a$ at time $t$ is
\beqa\label{Red2}
P_a (n,t) &=& {}_a\bra{n}\rhoh_a (t)\ket{n}_a\nn\\
          &=& e^{\frac{2}{\hbar}\mbox{Im}[f_0^A(t)+f_0^B(t)]}
          \,e^{|\mu_B\cos\theta-\mu_A\sin\theta|^2}\,\frac{|\mu_B\cos\theta+\mu_A\sin\theta|^{2n}}{n!},\nn\\
\eeqa
which is a Poisson distribution as expected from Eq. (\ref{gr7}).
\subsection{Time-evolution of an arbitrary state}\label{sec6}
An arbitrary initial state of the combined system is a linear combination of the tensor product states $\ket{n}_a\otimes\ket{m}_b$ where $\ket{n}_a\,(\ket{m}_b)$ is a number state in the Hilbert of the $a\,(b)$-oscillator. Since the evolution is linear, we only need to find the time evolution of the state $\ket{n}_a\otimes\ket{m}_b$. We have
\beqa\label{gen1}
\ket{\varphi (t)}=&& \hatU (t)\,\ket{n}_a\otimes\ket{m}_b,\nn\\
=&& \hatU_A (t)\otimes\hatU_B (t)\,\frac{(\hatad)^n}{\sqrt{n!}}\,\frac{(\hatbd)^m}{\sqrt{m!}}\,\ket{0},
\eeqa
by inserting the Bogoliubov transformations Eqs. (\ref{H2},\ref{H2-2}) into Eq. (\ref{gen1}), we find
\beqa\label{gen2}
\ket{\varphi (t)}= \frac{1}{\sqrt{n! m!}}\sum_{p=0}^n\sum_{q=0}^m && \binom{n}{p}\binom{m}{q}(-1)^{m-q}\,(\cos\theta)^{n-p+q}(\sin\theta)^{m-q+p}\nn\\
&& \times \hatU_A (t) (\hatAd)^{n+m-p-q}\otimes\hatU_B (t)(\hatBd)^{p+q}\ket{0}.
\eeqa
To proceed, we make use of the following identities
\beqa\label{gen3}
&& e^{-i \bar{f}_2^A\,\hatA/\hbar}\hatAd\,e^{i \bar{f}_2^A\,\hatA/\hbar}=\hatAd-\frac{i}{\hbar}\bar{f}_2^A,\nn\\
&& e^{-i f_1^A\,\hatAd\hatA/\hbar}\hatAd\,e^{i f_1^A\,\hatAd\hatA/\hbar}=e^{-i f_1^A /\hbar}\,\hatAd,
\eeqa
and rewrite Eq. (\ref{gen2}) as
\beqa\label{gen4}
\ket{\varphi (t)}=&& \frac{1}{\sqrt{n! m!}}\sum_{p=0}^n\sum_{q=0}^m \binom{n}{p}\binom{m}{q}(-1)^{m-q}\,(\cos\theta)^{n-p+q}(\sin\theta)^{m-q+p}\nn\\
&& \Big(e^{-i f_1^A/\hbar}\hatAd-\frac{i}{\hbar}\bar{f}_2^A\Big)^{n+m-p-q}\otimes\Big(e^{-i f_1^B/\hbar}\hatBd-\frac{i}{\hbar}\bar{f}_2^B\Big)^{p+q}\nn\\
&& \times\Big[\underbrace{\hatU_A (t)\otimes\hatU_B (t)\ket{0}}_{\mbox{see} \, Eq. (\ref{gr2})}\Big].
\eeqa
Therefore, in the coherent state basis $\{\ket{\alpha}_A\otimes\ket{\beta}_B \equiv\ket{\alpha}_A\ket{\beta}_B\}$, we find
\beqa\label{gen5}
{}_A\bra{\alpha}{}_B\bra{\beta}\varphi (t)\ra &=& \frac{(-1)^m}{\sqrt{n! m!}}\sum_{p=0}^n\sum_{q=0}^m \binom{n}{p}\binom{m}{q}(-1)^{q}\,(\cos\theta)^{n-p+q}(\sin\theta)^{m-q+p}\nn\\
&& \times\Big(e^{-i f_1^A/\hbar}\,\bar{\alpha}-\frac{i}{\hbar}\bar{f}_2^A\Big)^{n+m-p-q}\,\Big(e^{-i f_1^B/\hbar}\,\bar{\beta}-\frac{i}{\hbar}\bar{f}_2^B\Big)^{p+q}\nn\\
&& \times e^{-i(f_0^A (t)+f_0^B (t))/\hbar}\,e^{(|f_2^A (t)|^2+|f_2^B (t)|^2)/2\hbar^2}\nn\\
&& \times {}_A\la \alpha|{-\frac{i}{\hbar}f_2^A (t)\,e^{-i f_1^A (t)/\hbar}}\ra_A\,{}_B\la \beta|
             {-\frac{i}{\hbar}f_2^B (t)\,e^{-i f_1^B (t)/\hbar}}\ra_B.\nn\\
\eeqa
To obtain $\ket{\varphi(t)}$ in terms of the standard basis $\{\ket{n}_a\ket{m}_b\}$, we should replace $\hatU_A (t)\otimes\hatU_B (t)\ket{0}$ with Eq. (\ref{gr4}) and write $\hatAd\,(\hatBd)$ in terms of the $\hatad\,(\hatbd)$ using the inverse Bogoliubov transformations Eqs. (\ref{Inverse},\ref{Inverse-2}).
%
\section{Three interacting oscillators}\label{Seciii}
In this section, we consider the quantum dynamics of three interacting oscillators with equal frequencies $\omega_0$ coupling functions $g(t)$ and $g'(t)$, Fig. 3
\begin{figure}[h]
  \centering
  \includegraphics[width=.5\textwidth]{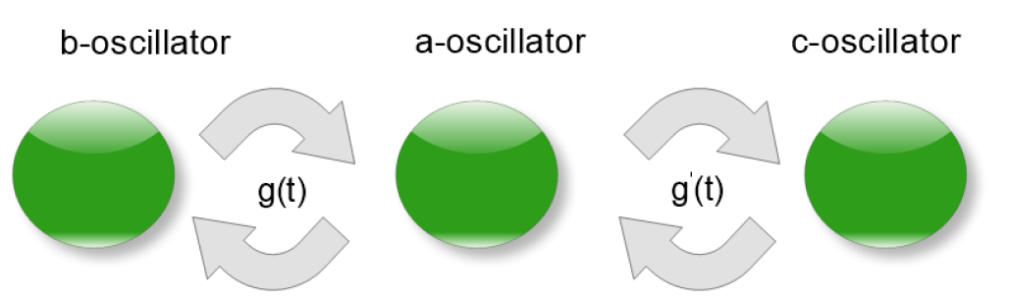}
  \caption{(color online) Three interacting oscillators with time-dependent coupling functions $g(t)$ and $g'(t)$. }\label{Fig3}
\end{figure}
The Hamiltonian of the system can be written as
\beq\label{H3}
\hatH (t)=\hbar\omega_0\,(\hatad\hata+\hatbd\hatb+\hatcd\hatc)
+\hbar g(t)\,(\hatad\hatb+\hata\hatbd)+\hbar g' (t) (\hatad\hatc+\hata\hatcd),
\eeq
which can also be written in matrix form
\beq\label{Matrix3}
\hatH (t)=\hbar\,\left(
                   \begin{array}{ccc}
                     \hata^\dag & \hatb^\dag & \hatc^\dag\\
                   \end{array}
                 \right)\underbrace{\left(
                          \begin{array}{ccc}
                            \omega_0 & g(t) & g'(t) \\
                            g(t) & \omega_0 & 0 \\
                            g'(t) & 0 & \omega_0 \\
                          \end{array}
                        \right)}_{\Lambda}\left(
                                 \begin{array}{c}
                                   \hata \\
                                   \hatb \\
                                   \hatc \\
                                 \end{array}
                               \right).
\eeq
By diagonalizing the coefficient matrix $\Lambda$ in Eq. (\ref{Matrix3}) the Hamiltonian can be diagonalized as
\beqa\label{Hdia}
\hatH (t)=\hbar(\omega_0-G(t))\,\hatA^\dag\hatA+\hbar\omega_0\,\hatB^\dag\hatB+
\hbar(\omega_0+G(t))\,\hatC^\dag\hatC,
\eeqa
where $G(t)=\sqrt{g^2 (t)+{g'}^2(t)}$ and the time-dependent energy levels in terms of the new modes $\hatA$, $\hatB$ and $\hatC$ are $E_A=\hbar(\omega_0-G(t))$, $E_B=\hbar\omega_0$ and $E_C=\hbar(\omega_0+G(t))$, respectively. The Bogoliubov transformations can be written using the orthogonal matrix $T$ ($TT^t=\mathrm{1}$) that diagonalizes the Hamiltonian
\beqa\label{Tmatrix}
&& \left(
  \begin{array}{c}
    \hatA \\
    \hatB \\
    \hatC \\
  \end{array}
\right)=\underbrace{\left(
          \begin{array}{ccc}
            -\frac{1}{\sqrt{2}} & \frac{g(t)}{\sqrt{2}G(t)} & \frac{{g'}(t)}{\sqrt{2}G(t)} \\
            0 & -\frac{{g'}(t)}{\sqrt{2}G(t)} & \frac{g(t)}{\sqrt{2}G(t)} \\
            \frac{1}{\sqrt{2}} & \frac{g(t)}{\sqrt{2}G(t)} & \frac{{g'}(t)}{\sqrt{2}G(t)} \\
          \end{array}
        \right)}_{T^t}\left(
                 \begin{array}{c}
                   \hata \\
                   \hatb \\
                   \hatc \\
                 \end{array}
               \right),\nn\\
&& \left(
  \begin{array}{c}
    \hata \\
    \hatb \\
    \hatc \\
  \end{array}
\right)=\underbrace{\left(
          \begin{array}{ccc}
            -\frac{1}{\sqrt{2}} & 0 & \frac{1}{\sqrt{2}} \\
            \frac{g(t)}{\sqrt{2}G(t)} & -\frac{{g'}(t)}{\sqrt{2}G(t)} & \frac{g(t)}{\sqrt{2}G(t)} \\
            \frac{{g'}(t)}{\sqrt{2}G(t)} & \frac{g(t)}{\sqrt{2}G(t)} & \frac{{g'}(t)}{\sqrt{2}G(t)} \\
          \end{array}
        \right)}_{T}\left(
                 \begin{array}{c}
                   \hatA \\
                   \hatB \\
                   \hatC \\
                 \end{array}
               \right),\nn\\
&& T^t\,\Lambda\,T=\left(
                     \begin{array}{ccc}
                       \omega_0-G(t) & 0 & 0 \\
                       0 & \omega_0 & 0 \\
                       0 & 0 & \omega_0+G(t) \\
                     \end{array}
                   \right).
\eeqa
%
\subsection{Evolution of the initial state}\label{sec7}
Let the initial state of the three oscillators defined by the bosonic modes ($\hata, \,\hatb, \,\hatc)$) be given by $|\psi(0)\ra=|1\ra_a\,|0\ra_b\,|0\ra_c$, then
\beqa\label{n1evo}
|\psi(0)\ra=&& \hata^\dag\,|0\ra_a\,|0\ra_b\,|0\ra_c,\nn\\
           =&& (-\frac{1}{\sqrt{2}}\hatA^\dag+\frac{1}{\sqrt{2}}\hatC^\dag)\,|0\ra_A\,|0\ra_B\,|0\ra_C,\nn\\
           =&& -\frac{1}{\sqrt{2}}|1\ra_A\,|0\ra_B\,|0\ra_C+\frac{1}{\sqrt{2}}|0\ra_A\,|0\ra_B\,|1\ra_C.
\eeqa
The evolution operator corresponding to the Hamiltonian Eq. (\ref{Hdia}) is
\beqa\label{U3}
\hat{U}(t)=&& e^{-if_{-} (t)\,\hatA^\dag\hatA}\,e^{-i\omega_0 t\,\hatB^\dag\hatB}
\,e^{-if_{+} (t)\,\hatC^\dag\hatC},
\eeqa
where
\beqa
f_{\pm}(t)=&& \omega_0 t\pm\int_0^t dt'\,G(t'),\nn\\
          =&& \omega_0 t\pm\,\eta(t).
\eeqa
Therefore,
\beqa\label{psi3t}
|\psi(t)\ra=&& -\frac{1}{\sqrt{2}}e^{-if_{-}(t)}|1\ra_A\,|0\ra_B\,|0\ra_C+\frac{1}{\sqrt{2}}
e^{-if_{-}(t)}|0\ra_A\,|0\ra_B\,|1\ra_C,\nn\\
           =&&  \Bigg[-\frac{1}{\sqrt{2}}e^{-if_{-} (t)}
\bigg(-\frac{1}{\sqrt{2}}\hata^\dag+\frac{g(t)}{\sqrt{2}G(t)}\hatb^\dag+
\frac{g'(t)}{\sqrt{2}G(t)}\hatc^\dag\bigg)\nn\\
            && +
\frac{1}{\sqrt{2}}e^{-if_{+}(t)}\bigg(\frac{1}{\sqrt{2}}\hata^\dag+
\frac{g(t)}{\sqrt{2}G(t)}\hatb^\dag+\frac{g'(t)}
{\sqrt{2}G(t)}\hatc^\dag\bigg)\Bigg]\,|0\ra_A\,|0\ra_B\,|0\ra_C,\nn\\
           =&& \Bigg[\frac{1}{2}\bigg(e^{-if_{+}(t)}+e^{if_{-}(t)}\bigg)\hata^\dag+
\frac{g(t)}{2G(t)}\bigg(e^{-if_{+}(t)}-e^{if_{-}(t)}\bigg)\hatb^\dag
\nn\\
&& +\frac{g'(t)}{2G(t)}\bigg(e^{-if_{+}(t)}-e^{if_{-}(t)}\bigg)\hatc^\dag\Bigg]\,|0\ra_{ABC},\nn\\
           =&& e^{-i\omega_0 t}\cos\eta(t)\,|1\ra_a\,|0\ra_b\,|0\ra_c-
\frac{ig(t)}{G(t)}e^{-i\omega_0 t}\sin\eta(t)\,|0\ra_a\,|1\ra_b\,|0\ra_c
            \nn\\
            && -\frac{ig'(t)}{G(t)}e^{-i\omega_0 t}\sin\eta(t)\,|0\ra_a\,|1\ra_b\,|1\ra_c.
\eeqa
From Eq. (\ref{psi3t}), we find that the probability that the state remains at the initial state is
$P_{100}(t)=\cos^2\eta(t)$ and the probability that the excitation be transferred to $b$-oscillator or $c$-oscillator is given by $P_{010}(t)=(g(t)/G(t))^2\,\sin^2\eta(t)$ and $P_{001}(t)=(g'(t)/G(t))^2\,\sin^2\eta(t)$ respectively.

For the initial state $|\psi(0)\ra=|2\ra_a|0\ra_b|0\ra_c$ we will find the evolved state as
\beqa\label{psi3t2}
|\psi(t)\ra &=& \frac{1}{2}e^{-2i\omega_0 t}(1+\cos(2\eta(t)))\,|2\ra_a|0\ra_b|0\ra_c\nn\\
&&+ \frac{g^2(t)}{2G^2(t)}e^{-2i\omega_0 t}(\cos(2\eta(t))-1)\,|0\ra_a|2\ra_b|0\ra_c\nn\\
&&+ \frac{g'^2(t)}{2G^2(t)}e^{-2i\omega_0 t}(\cos(2\eta(t))-1)\,|0\ra_a|0\ra_b|2\ra_c\nn\\
&&+ \frac{g(t)}{\sqrt{2}G(t)}e^{-2i\omega_0 t}(-i\sin(2\eta(t)))\,|1\ra_a|1\ra_b|0\ra_c\nn\\
&&+ \frac{g'(t)}{\sqrt{2}G(t)}e^{-2i\omega_0 t}(-i\sin(2\eta(t)))\,|1\ra_a|0\ra_b|1\ra_c\nn\\
&&+ \frac{g(t)g'(t)}{\sqrt{2}G^2(t)}e^{-2i\omega_0 t}(\cos(2\eta(t))-1)\,|0\ra_a|1\ra_b|1\ra_.
\eeqa
One can easily deduce that (i) the probability that at time $t$ the excitation be transferred from $a$-oscillator to the right or $c$-oscillator is $P_{a\rightarrow c}=(g'(t)/G(t))^2\sin^2\eta(t)$ (ii) the probability that at time $t$ the excitation be transferred from $a$-oscillator to the left or $b$-oscillator is $P_{a\rightarrow b}=(g(t)/G(t))^2\sin^2\eta(t)$ (iii) the probability that at time $t$ the excitation remains with the $a$-oscillator is $P_{a\rightarrow a}=\cos^2\eta(t)$. Therefore, for the initial state given by
$|\psi(0)\ra=|n\ra_a|1\ra_b|0\ra_c$ the probability that at time $t$ we find the evolved state $|\psi(t)\ra$ at $|i\ra_a|j\ra_b|k\ra_c$ is
\beqa\label{cijk}
P_{ijk}(t)=&& P_{\{|n\ra_a|0\ra_b|0\ra_c\rightarrow |i\ra_a|j\ra_b|k\ra_c\}}(t)\nn\\
          =&& \binom{n}{i,j,k} \,\cos^{2i}\eta(t)\,\Big(\frac{g^2(t)}{G^2(t)}\sin^2\eta(t)\Big)^j\,
             \Big(\frac{{g'}^2(t)}{G^2(t)}\sin^2\eta(t)\Big)^k,
\eeqa
where $\binom{n}{i,j,k}=\frac{n!}{i! j! k!},\,\,(i+j+k=n),$ and $\sum\limits_{{i+j+k=n}} P_{ijk}(t)=1$, as expected.
\begin{figure}[h]
  \centering
  \includegraphics[width=.55\textwidth]{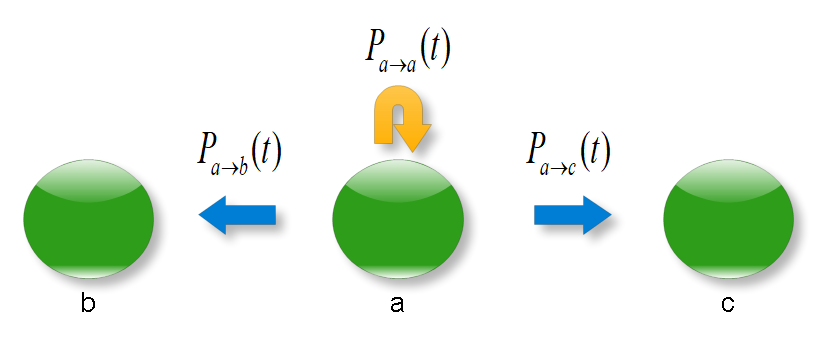}
  \caption{(color online) The transition probabilities for three interacting oscillators with time-dependent coupling functions $g(t)$ and $g'(t)$. }\label{Fig4}
\end{figure}
%
\subsection{$n$-interacting oscillators}\label{Secn}
%
\begin{figure}[h]
  \centering
  \includegraphics[width=.8\textwidth]{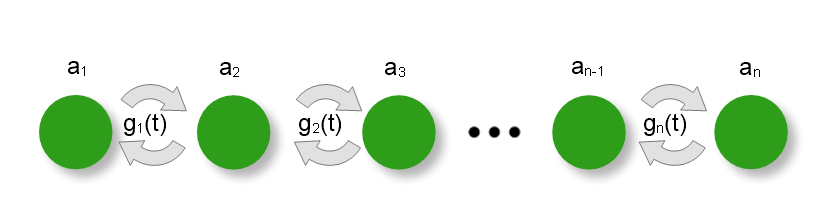}
  \caption{(color online) Linearly-interacting oscillators with time-dependent coupling functions $g_k(t),\,\,(k=1,2,\cdots,n-1)$.} \label{Fig5}
\end{figure}
For $n$ linearly-interacting oscillators with equal frequencies $\omega_0$, the Hamiltonian is
\beq\label{Hn}
\hatH (t)=\sum_{k=1}^n \hbar\omega_0\,\hata^\dag_k \hata_k+\sum_{k=1}^{n-1} \hbar g_k (t)\,
(\hata_k \hata^\dag_{k+1}+\hata^\dag_{k}\hata_{k+1}),
\eeq
which can also be written in matrix form as
\beqa\label{matn}
&& \hatH (t)=\nn\\
&& \hbar\left(
                 \begin{array}{cccc}
                   \hata^\dag_1 & \hata^\dag_2 & \cdots & \hata^\dag_n \\
                 \end{array}
               \right)
\underbrace{\left(
            \begin{array}{ccccc}
              \omega_0 & g_1 (t) & 0 & 0 & \cdots \\
              g_1 (t) & \omega_0 & g_2 (t) & 0 & \cdots \\
              0 & g_2(t) & \omega_0 & g_3 (t) & \vdots \\
              \vdots & \vdots & \vdots & \ddots & g_{n-1} (t) \\
              0 & 0 & \cdots & g_{n-1} (t) & \omega_0 \\
            \end{array}
          \right)}_{\Gamma}\left(
                   \begin{array}{c}
                     \hata_1 \\
                     \hata_2 \\
                     \vdots \\
                     \hata_1 \\
                   \end{array}
                 \right).\nn\\
\eeqa
Now by diagonalizing the coefficient matrix $\Gamma$ we can obtain the corresponding Bogoliubov transformations to diagonalize the Hamiltonian and follow the same steps we took for the three interacting oscillators.
\section{Conclusion}\label{sec7}
The main result obtained in Sec. (\ref{Secii}) was Eq. (\ref{H12}) describing the probability of finding the oscillators in the state $|n-s\ra_a\otimes|s\ra_b$ at time $t$ if the initial state was $|n\ra_a\otimes|0\ra_b$. The generated state Eq. (\ref{psit}) was a binomial state and the probability given in Eq. (\ref{H12}) was like a classical one-dimensional random walk with time-dependent right ($p(t)$) and left ($q(t)$) probabilities depending on detuning $\triangle$ and coupling function $g(t)$. For the initial state $|\alpha\ra_a\otimes|s\ra_b$, the evolved state was given in Eq. (\ref{coh1}) and in this case the probability of finding the oscillators in the state $|n-s\ra_a\otimes|s\ra_b$ is
\beqa
\underbrace{\frac{e^{-|\alpha|^2}|\alpha\cos^2 (\theta)[1+\tan^2(\theta)e^{i\delta f(t)}]|^{2n}}{n!}}_{Poisson}\underbrace{\binom{n}{s}\Bigg|\frac{\tan(\theta)[e^{i\delta f(t)}-1]}{1+\tan^2(\theta)e^{i\delta f(t)}}\Bigg|^2}_{binomial},\nn
\eeqa
which is the product of a Poisson and a binomial distribution.

In the following we assumed an external source acting on the $b$-oscillator and found the evolved state Eq. (\ref{gen4}) at time $t$ for the initial state $|n\ra_a\otimes|m\ra_b$. For the special case $|0\ra_a\otimes|0\ra_b$, the generated state was the product of two coherent states given in Eq. (\ref{gr7}).

For three linearly-interacting oscillators with time-dependent coupling functions $g_1 (t)$ and $g_2(t)$, we found the exact propagator Eq. (\ref{U3}) and calculated the probability transition $P_{a\rightarrow a}(t)$ that the excitation remains with the $a$-oscillator or be transferred to $b(c)$-oscillator with probability transitions $P_{a\rightarrow b(c)}(t)$ respectively.

In the following, based on the results obtained from two and three-linearly interacting oscillators, we could generalize the problem to $n$-linearly interacting oscillators with time-dependent coupling functions $g_1 (t), g_2 (t), \cdots, g_n (t)$. In this case, one needs to diagonalize the coefficient matrix $\Gamma$ in Eq. (\ref{matn}) to find the corresponding Bogoliubov transformations and follow the same steps we took for two or three-linearly interacting oscillators.
%
%
	
%

%
%
\end{document}